\documentclass[journal]{IEEEtran}

\ifCLASSINFOpdf
\else
   \usepackage[dvips]{graphicx}
\fi
\usepackage{url}

\usepackage{graphicx}
\usepackage{subfigure}
\usepackage{makecell}
\usepackage{amsmath}

\usepackage[pscoord]{eso-pic}
\newcommand{\placetextbox}[3]{
 \setbox0=\hbox{#3}
 \AddToShipoutPictureFG*{ \put(\LenToUnit{#1\paperwidth},\LenToUnit{#2\paperheight}){\vtop{{\null}\makebox[0pt][c]{#3}}}
 }
 }
 \placetextbox{.5}{0.045}{\small{1545-598X~\copyright~2020 IEEE. Personal use is permitted, but republication/redistribution requires IEEE permission.}}
 \placetextbox{.5}{0.030}{\small{See https://www.ieee.org/publications/rights/index.html for more information.}}
 


\begin{document}

\title{An Unsupervised Generative Neural Approach for InSAR Phase Filtering and Coherence Estimation}

\IEEEaftertitletext{\vspace{-1\baselineskip}}

\author{Subhayan Mukherjee, Aaron Zimmer, Xinyao Sun, Parwant Ghuman, Irene Cheng \IEEEmembership{Senior Member, IEEE}

\thanks{Manuscript submitted January 31, 2020; resubmitted April 30, 2020; revised June 9, 2020 and July 13, 2020; accepted July 15, 2020. Research supported by NSERC Discovery Grant RGPIN-2018-04367 and Department of National Defence/NSERC Discovery Grants Supplements DGDND-2018-00020.}
\thanks{S. Mukherjee, X. Sun and I. Cheng are with the Department of Computing Science, University of Alberta, Edmonton, AB T6G 2R3 Canada (e-mails: mukherje,xinyao1,locheng@ualberta.ca).}
\thanks{A. Zimmer and P. Ghuman are with 3vGeomatics, Vancouver, BC V5Y 0M6 Canada (e-mails: azimmer,pghuman@3vgeomatics.com).}
\thanks{Digital Object Identifier 10.1109/LGRS.2020.3010504}}

\markboth{IEEE GEOSCIENCE AND REMOTE SENSING LETTERS has accepted this article for publication in a future issue. Content (except pagination) is final as presented.}
{Shell \MakeLowercase{\textit{et al.}}: Bare Demo of IEEEtran.cls for IEEE Journals}
\maketitle

\begin{abstract}
Phase filtering and pixel quality (coherence) estimation is critical in producing Digital Elevation Models (DEMs) from Interferometric Synthetic Aperture Radar (InSAR) images, as it removes spatial inconsistencies (residues) and immensely improves the subsequent unwrapping. Large amount of InSAR data facilitates Wide Area Monitoring (WAM) over geographical regions. Advances in parallel computing have accelerated Convolutional Neural Networks (CNNs), giving them advantages over human performance on visual pattern recognition, which makes CNNs a good choice for WAM. Nevertheless, this research is largely unexplored. We thus propose ``GenInSAR'', a CNN-based generative model for joint phase filtering and coherence estimation, that directly learns the InSAR data distribution. GenInSAR's unsupervised training on satellite and simulated noisy InSAR images outperforms other five related methods in total residue reduction (over 16$\frac{1}{2}$\% better on average) with less over-smoothing/artefacts around branch cuts. GenInSAR's Phase, and Coherence Root-Mean-Squared-Error and Phase Cosine Error have average improvements of 0.54, 0.07, and 0.05 respectively compared to the related methods.
\end{abstract}


\IEEEpeerreviewmaketitle

\section{Introduction}

\IEEEPARstart{I}{nSAR} or Interferometric Synthetic Aperture Radar is an emerging, highly successful remote sensing technique for measuring several geophysical quantities like surface deformation \cite{paperD}. It is based on generating an interferogram as the complex difference of two SAR acquisitions of the same scene from slightly different view angles. The wrapped interferometric phase is then unwrapped to subsequently produce a Digital Elevation Model (DEM). However, several decorrelation factors create strong phase noise, affecting unwrapping and DEM accuracy \cite{paperM}. Thus phase filtering is preferred, even when it results in some decrease in resolution and increase in spatial correlation \cite{paperL} and we need filters adapted to enhance phase rather than amplitude \cite{paperC}. Filtering the real and imaginary parts of the complex phase in its wrapped form can avoid blurring edges \cite{paperH,paperF}, whereas unwrapping before filtering increases computation and potentially decreases accuracy \cite{paperD}. Due to the non-stationary nature of InSAR signals, simple boxcar averaging and non-adaptive filtering methods tend to distort the phase \cite{paperL,paperH}. Methods that adapt their parameters based on, e.g. local phase quality (coherence) yield better results, as coherence is related to phase noise deviation \cite{paperM,paperF}. Early spatial methods like Lee \cite{lee98}, frequency based methods like Goldstein \cite{goldstein98} and their numerous improvements \cite{insarmet6,insarmet7,insarmet8,insarmet9,insarmet10} adapt to the local fringe direction and/or local noise. Frequency based methods gradually evolved into the wavelet domain \cite{insarmet12,insarmet13,insarmet14} to simplify the separation of true phase from noise \cite{paperM} but struggled to filter dense fringes, whereas spatial methods in general sacrificed spatial resolution \cite{paperI}. The additive noise model of interferometric phase \cite{lee98} inspired early filtering methods which assumed a stationary and consistent phase over the filtering window, but real-world challenges of strong topographic changes and restrictions imposed on window size motivated more recent non-linear models \cite{paperB} and per-pixel filtering \cite{perpix2016}. Recent advances in parallel computing architectures have motivated parallelism in the InSAR processing pipeline \cite{parallel2015}, which is critical to our proposed phase filtering method (``GenInSAR'') for InSAR-based Wide Area Monitoring (WAM) across geographical regions involving petabytes of data. Thus, we use a Convolutional Neural Network (CNN) architecture which seamlessly integrates with modern parallel architectures built on Graphics Processing Units (GPUs) to outperform humans on pattern recognition tasks. Use of CNNs in InSAR phase processing has been limited to volcano deformation monitoring \cite{paperE} via transfer learning from optical images, but not training directly on InSAR data. Recent CNN-based InSAR phase filtering and coherence estimation/classification \cite{ieeesp,ieeesc} trained directly on InSAR data, but with separate CNNs for filtering and coherence estimation, and ``raw'' coherence generated/preprocessed using traditional methods. In contrast, our GenInSAR filters phase and estimates coherence jointly, using a single neural network, and predicts the center pixel's distribution given only its neighborhood. It is thus ``embarrassingly parallel'' \cite{embpar2012}, whereas non-local filters \cite{nlinsar,nlsar} require computing patch similarity and suffer from terrace-like DEM artefacts, over-smoothing and ``rare patch'' effect \cite{paperG}. For similar computational concerns, we avoid strategies that are iterative \cite{paperL,paperC}, multi-stage \cite{paperG,paperC} and require optimization during inference, e.g. via sparse coding \cite{paperD}. Iterative strategies can also result in loss of detailed features \cite{paperL}.

We propose a novel InSAR phase filter inspired by Mixture Density Networks (MDN) \cite{Bishop94mdn}. Our CNN convolutional layers operate on their input (phase) receptive field, and help predict the parameters $(\vec{\mu},\vec{\sigma})$ of a bi-variate (\textbf{R}eal, \textbf{I}maginary) Gaussian distribution of the center pixel: filtered pixel $\vec{\mu}=(\mu_{\textbf{R}},\mu_{\textbf{I}})$, having coherence $\gamma$ as a function of $\vec{\sigma}=(\sigma_{\textbf{R}},\sigma_{\textbf{I}})$ (Eq. \ref{eq:gamma}).

Our approach, GenInSAR facilitates learning the real InSAR data distribution from huge datasets generated by the rapidly increasing use of SAR satellites. Being a generative model, sampling from this distribution generates new interferograms which are slight variations of the filter output, and can be utilized to improve the InSAR pipeline. GenInSAR produces state-of-art results surpassing NLInSAR \cite{nlinsar}, without being prohibitively slow for most production situations like WAM.

\section{Proposed Method}
\label{sec:method}
\begin{figure}[h]
\centering
\includegraphics[scale=0.6]{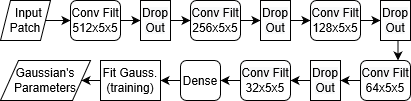}
\caption{Architecture of our proposed method GenInSAR.} 
\label{fig:arch}
\end{figure}

Fig. \ref{fig:arch} shows GenInSAR's architecture. Its input is a $21\times21$ phase patch centered around the pixel to be filtered. That pixel is zeroed out in the input patch to avoid learning the identity mapping. Thus, GenInSAR learns to ignore it during prediction. Significantly smaller patch sizes like $11\times11$ reduce the receptive field, resulting in loss of details in the filtered phase and increasing bias in the predicted coherence. We can understand this more clearly in the training (fitting) and testing (prediction) steps. While training, patches extracted from a fixed set of phase images (training set) are input to the model. We set 50\% dropout rate \cite{dropout2014} during training to prevent over-fitting. Dropout randomly sets 50\% activations of its preceding layer to zero. Intuitively, this forces the network to learn simpler mappings for each training example, thus preventing over-fitting. Convolutional layers \cite{cnn_orig1998} of fixed filter size $5\times5$ but decreasing filter counts (512, 256, 128, 64, 32), with each followed by Exponential Linear Unit activation \cite{elu_2016} (not shown) with $\alpha=1.0$ promote fast convergence and non-linear mappings. It also allows negative outputs $\geq -1.0$ (lower limit of $\cos \theta$ and $\sin \theta$). Specifically, we use depth-wise separable 2D convolutions \cite{depsepconv2d} with one filter per input channel (depth) $\hat{G}_{k, l, m}=\sum_{i, j} \hat{K}_{i, j, m} \cdot F_{k+i-1, l+j-1, m}$ for fast computation and convergence, where the $m^{th}$ kernel $\hat{K}_{m}$ is applied to $m^{th}$ $k \times l$ input feature map $F_{m}$ to obtain $m^{th}$ output feature map $\hat{G}_{m}$, followed by a $1\times1$ convolution to combine the outputs. Finally, following MDN working principle, dense connections (weighted sums of all filter outputs) to the distribution fitting module outputs those Gaussian parameter values ($\vec{\mu},\vec{\sigma}$) for the real and imaginary channel which maximizes the input patch's central pixel (training target) likelihood.

Thus, our training is completely unsupervised, as we learn from the input data itself, without requiring its ``clean'' version as the training target. The central pixel ${t}^{q}$ (surrounded by its neighborhood pixels, ${x}^{q}$) is treated as a sample drawn from the reference Gaussian distribution chosen to best encompass all $n$ training set examples $\left\{{x}^{q},{t}^{q}\right\}$, by minimizing the loss $E=-\ln \prod_{q=1}^{n} p\left({t}^{q} | {x}^{q}\right) p\left({x}^{q}\right)$ via gradient descent back-propagation using Adam optimizer \cite{adam2015}. The network is thus trained to parameterize a Gaussian density that best encompasses observed (noisy) data, by minimizing $E$. During testing, dropout and distribution fitting (optimization) are not required.

GenInSAR does not train to predict $\gamma$. It is directly computed from the predicted $\vec{\sigma}$. A nice property of $\gamma$ is that it seems to be a better measure of filtering quality and filter output reliability, which partially depends on the spatial noise pattern (neighborhood), not just the noise underlying the center pixel. Considering two SAR acquisitions $(u_{1},u_{2})$ with resulting interferometric unwrapped phase $\theta$ having probability density $p(\theta)$, variance $\sigma_{\theta}^{2}$, and real and imaginary components $(R,I)$ with predicted variances $(\sigma_{R}^2,\sigma_{I}^2)$, we approximate $\gamma$ as the normalized index of mutual linear predictability of random variables $u_{1}$ and $u_{2}$, thus quantifying noise in interferometric acquisitions:
$\gamma=\frac{E\left\{u_{1} u_{2}^{*}\right\}}{\sqrt{E\left\{\left|u_{1}\right|^{2}\right\}} \sqrt{E\left\{\left|u_{2}\right|^{2}\right\}}}$
$=E\left\{u_{1}u_{2}^{*}\right\}$
since for our normalized input phasors, $\gamma$'s denominator reduces to 1.
\begin{flushleft}
$\gamma=\int_{-\infty}^{\infty} p(\theta) e^{i \theta} d \theta$
$=\int_{-\infty}^{\infty} \frac{e^{-\frac{\theta^{2}}{2\sigma_{\theta}^{2}}} }{\sigma_{\theta} \sqrt{2 \pi}}[\cos \theta+i \sin \theta] d \theta$
$=e^{-\frac{\sigma_{\theta}^{2}}{2}}$
\end{flushleft}
\begin{flushleft}
Again, $\sigma_{R}^{2}+\sigma_{I}^{2}$
$=E\left\{R^{2}\right\}-E\{R\}^{2}+E\left\{I^{2}\right\}-E\{I\}^{2}$
\end{flushleft}
\begin{flushleft}
$=\int_{-\infty}^{\infty} p(\theta) \cos ^{2} \theta d \theta + \int_{-\infty}^{\infty} p(\theta) \sin ^{2} \theta d \theta$
\end{flushleft}
\begin{flushright}
$-\left(\int_{-\infty}^{\infty} p(\theta) \cos \theta d \theta\right)^{2}-\left(\int_{-\infty}^{\infty} p(\theta) \sin \theta d \theta\right)^{2}$
\end{flushright}
$=e^{-\sigma_{\theta}^{2}}\left(\cosh \left(\sigma_{\theta}^{2}\right)+\sinh \left(\sigma_{\theta}^{2}\right)\right)-e^{-\sigma_{\theta}^{2}}-0$
\begin{flalign}
\label{eq:gamma}
=1-e^{-\sigma_{\theta}^{2}}
=1-\gamma^{2}
\implies \gamma = \sqrt{1 - (\sigma_{R}^2 + \sigma_{I}^2)} &&
\end{flalign}

GenInSAR trains on patches to best utilize the training data, but prediction on patches is slow, especially for large patch sizes. To solve this, we split the trained model into Convolver (all Conv layers) $+$ Combiner (Dense layer onward). During training, Convolver does not use padding, but we pad during prediction to ensure that the output and input image sizes match up along image borders. From Fig. \ref{fig:arch}, we can infer that the Convolver outputs a 32-channel pixel for each input image pixel, which the Combiner uses to output that pixel's $(\vec{\mu},\vec{\sigma})$. Since the Combiner operates on individual pixels, we use a very large batch size for the Combiner during prediction (4096, limited only by GPU memory) for high time efficiency.

We assume a Gaussian distribution for the unwrapped phase noise, to approximate the InSAR multiplicative speckle noise distribution. Generally, a Gaussian Mixture Model can approximate any distribution arbitrarily well by adding more terms, but we lower the free parameter count to achieve a lower bias. Moreover, since the number of effective samples is low, highly accurate characterization of the true underlying distribution is not necessary, since the mean for the two distributions will be equivalent and the only difference will be in the variance. Thus, the variance might be slightly underestimated, but this is a common problem with most coherence estimators.

GenInSAR's main advantage lies in learning the distribution of real data by directly training on it, without relying on assumptions. This is very useful, as countless images are being acquired daily by an ever-increasing number of SAR satellites around the world, creating a huge archive of real training data. Although most traditional filters rely on the spatial context to estimate the phase and coherence, GenInSAR's coherence qualitatively appears to be more of an indication of the confidence of its estimated phase. This is a useful feature, because in real-world machine learning applications, GenInSAR might encounter a feature that it was not trained on, so there might be some error in its phase estimation. In that case, it would predict a low coherence, and that pixel will be down weighted or removed in subsequent stages of InSAR processing. Additionally, in areas that are noisy but very smooth, the predicted coherence might be biased up slightly because GenInSAR is more confident of its estimated phase, as it made better use of the contextual information.


\section{Results and Discussion}

GenInSAR and the CNN-based InSAR filter mentioned earlier (hereafter referred to as ``CNN-InSAR'' \cite{ieeesp}) were implemented in Keras \cite{keras} (Tensorflow-GPU backend). Boxcar, Goldstein \cite{goldstein98} (outputs only phase), NLInSAR \cite{nlinsar}, and NLSAR \cite{nlsar} were implemented in OpenCL 1.2. All methods were executed on a 8 GB NVIDIA 1070 GPU. We present the qualitative and quantitative results of those experiments for real and simulated images respectively. The metrics used for quantitative analysis are Root-Mean-Square-Error (RMSE) of the InSAR phase, and coherence, Residue Reduction Percentage (RRP) \cite{paperL,paperM,paperI}, and Phase Cosine Error, $\epsilon_{cos}^{\Delta\theta}$ (Eq. \ref{eq:pce}) where $g_{i}$ and $\bar{f_{i}}$ denote $i^{th}$ ground truth and complex conjugate of filtered pixels of an  $n$ pixel interferogram. Residues are phase inconsistencies emphasized by computing curl of phase differences over the range of a reduced closed integral loop of four spatially adjacent pixels \cite{paperA,paperI}, which are non-zero if residues are present. Most residues are caused by noise. However, few arise from signal structure, like steep change in topography or heavy deformations, and those residues should be preserved during filtering. Filtering should remove all other residues to facilitate phase unwrapping. Those that cannot be removed should have low values in the filter's output coherence map; this prevents error propagation during phase difference integration by the unwrapper. Hence, filtering aims to reduce residues (high RRP) but preserve details (low Phase RMSE, $\epsilon_{cos}^{\Delta\theta}$). These criteria drive our evaluations:

\subsubsection{Experiments using satellite InSAR images}

\begin{figure*}
\centering

\subfigure[Input Phase]{\label{fig:qual_inphase}\includegraphics[width=42mm]{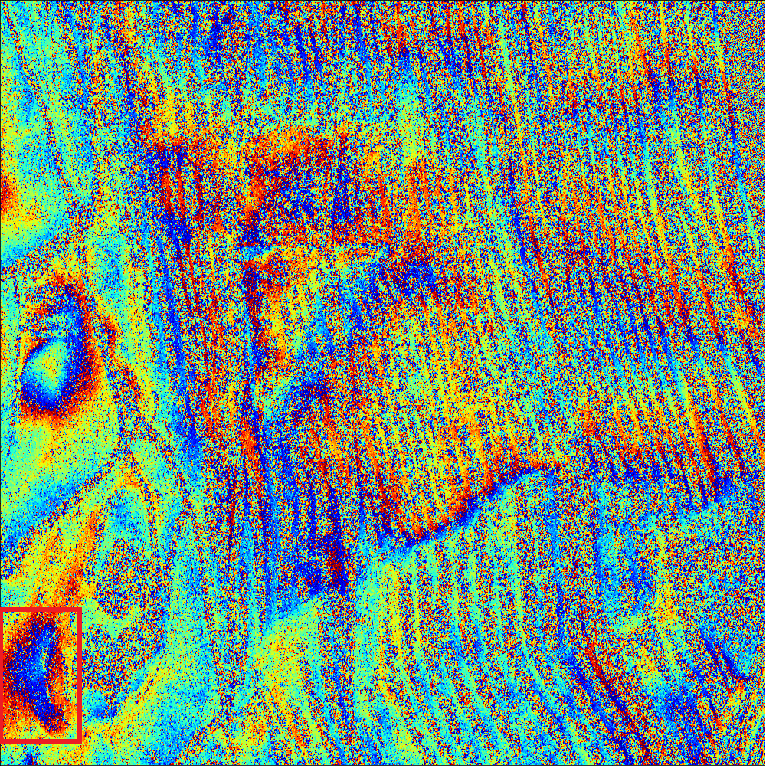}}
\enskip
\subfigure[Goldstein Phase]{\label{fig:gold_phase}\includegraphics[width=42mm]{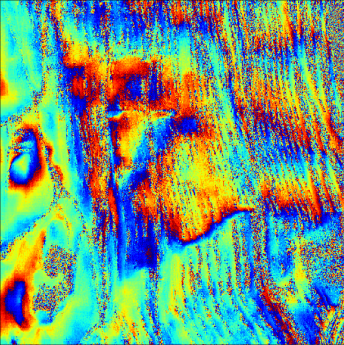}}
\enskip
\subfigure[Boxcar Phase]{\label{fig:box_phase}\includegraphics[width=42mm]{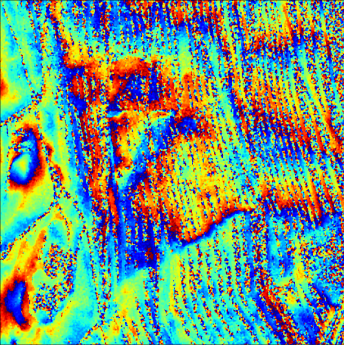}}
\enskip
\subfigure[Boxcar Coherence]{\label{fig:box_coh}\includegraphics[width=42mm]{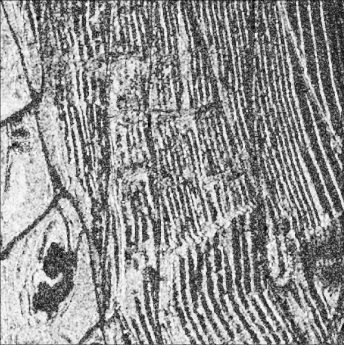}}

\subfigure[NLInSAR Phase]{\label{fig:nlinsar_phase}\includegraphics[width=42mm]{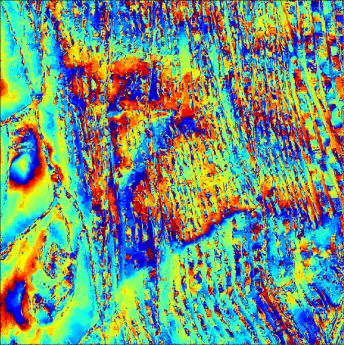}}
\enskip
\subfigure[NLInSAR Coherence]{\label{fig:nlinsar_coh}\includegraphics[width=42mm]{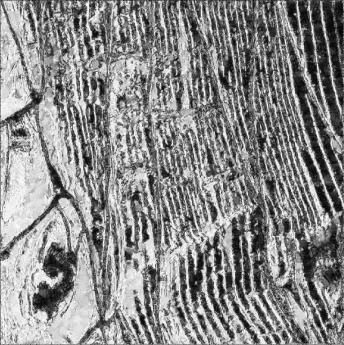}}
\enskip
\subfigure[NLSAR Phase]{\label{fig:nlsar_phase}\includegraphics[width=42mm]{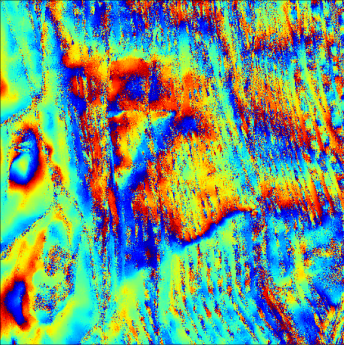}}
\enskip
\subfigure[NLSAR Coherence]{\label{fig:nlsar_coh}\includegraphics[width=42mm]{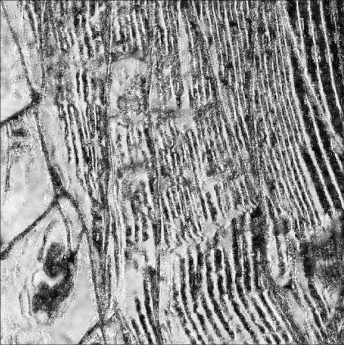}}

\subfigure[CNN-InSAR Phase]{\label{fig:cnninsar_phase}\includegraphics[width=42mm]{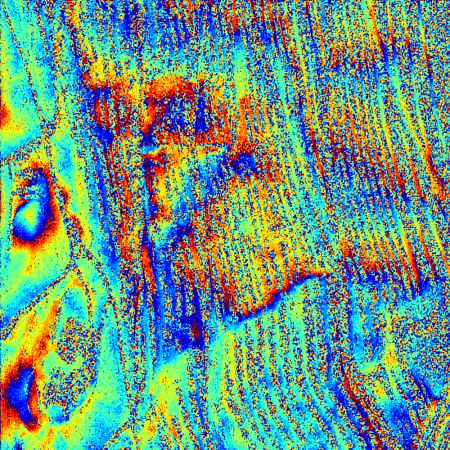}}
\enskip
\subfigure[CNN-InSAR Coherence]{\label{fig:cnninsar_coh}\includegraphics[width=42mm]{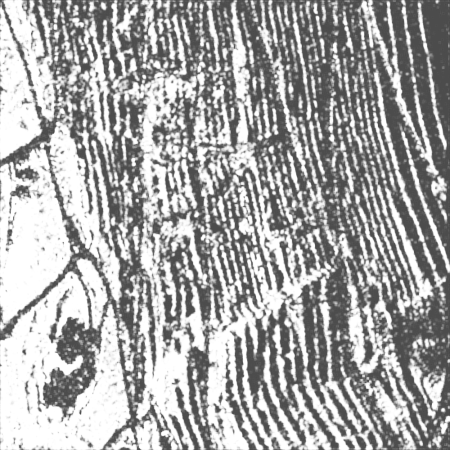}}
\enskip
\subfigure[Proposed (GenInSAR) Phase]{\label{fig:proposed_phase}\includegraphics[width=42mm]{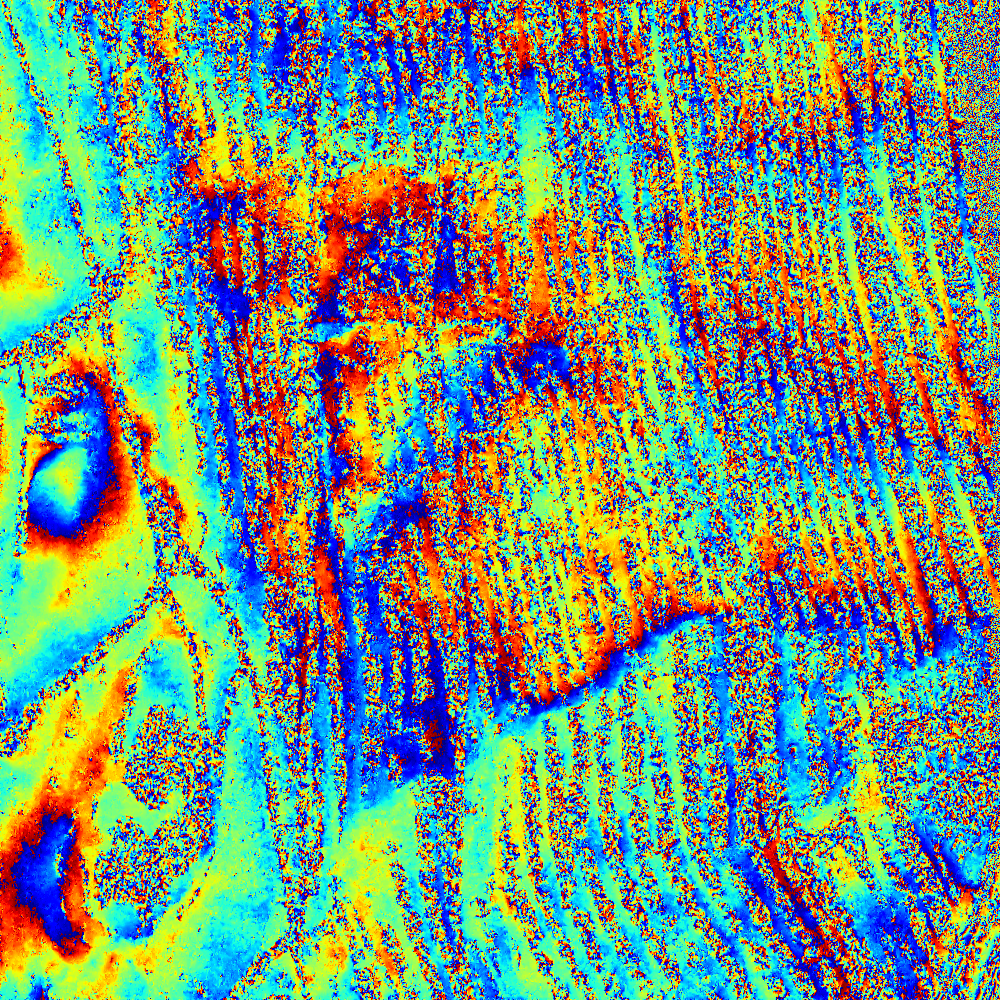}}
\enskip
\subfigure[Proposed (GenInSAR) Coherence]{\label{fig:proposed_coh}\includegraphics[width=42mm]{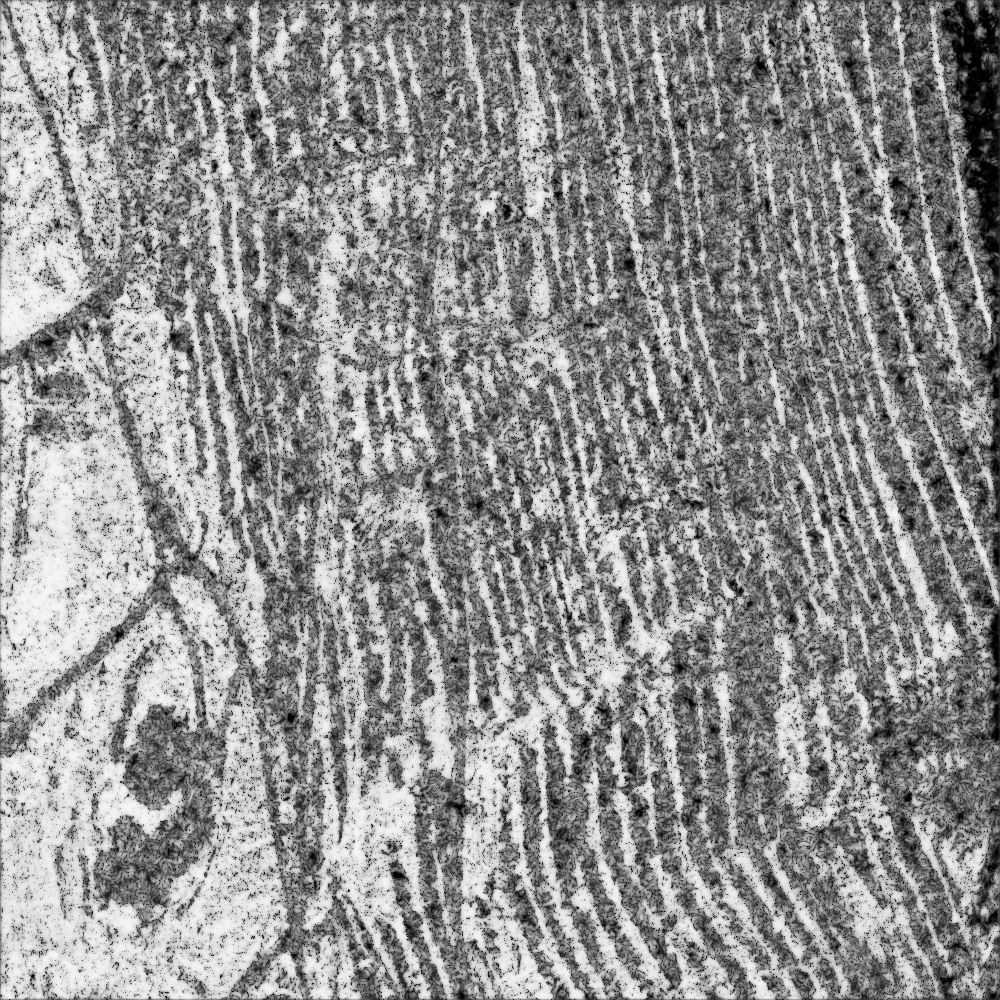}}

\caption{Filtered phase and coherence outputs for satellite InSAR images from GenInSAR (total residue reduction, less over-smoothing/artefacts around branch cuts) and five existing methods. Phase and coherence are coloured between --$\pi$ (blue) to +$\pi$ (red), and 0 (black: low) to 1 (white: high) respectively.}
\label{fig:qualcomp}
\end{figure*}

\begin{figure}
\centering

\includegraphics[width=15mm]{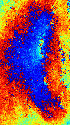}
\enskip
\includegraphics[width=15mm]{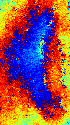}
\enskip
\includegraphics[width=15mm]{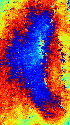}
\enskip
\includegraphics[width=15mm]{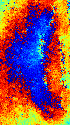}
\enskip
\includegraphics[width=15mm]{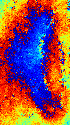}

\caption{GenInSAR outputs (fine, pixel-wise differences) obtained by sampling the Gaussian predicted for each pixel of the same noisy input (Fig. \ref{fig:qual_inphase}, red square). It is useful in data augmentation and testing InSAR processing chains / error analysis. Phase visualizations coloured from --$\pi$ (blue) to +$\pi$ (red).}
\label{fig:genifg}
\end{figure}

GenInSAR was trained for 100 epochs on 5 million $21\times21$ patches (in batches of 64 patches) extracted from 300 $1000\times1000$ pixels interferograms obtained from several different sensors at different resolutions. We tested the trained model on $1000\times1000$ pixels interferograms of a mining site. Fig. \ref{fig:qualcomp} shows outputs for a test interferogram using GenInSAR and other methods. GenInSAR is a generative model: its filtered output per pixel $(i,j)$ corresponds to the mean $\vec{\mu_{i,j}}$ (and $\vec{\sigma_{i,j}}$) of the distribution predicted for that pixel. To show this, we randomly sample five times from the normal distribution $N(\vec{\mu_{i,j}},\vec{\sigma_{i,j}}*\alpha)$ setting $\alpha=0.1$, and generate pixel $\vec{P}_{i,j}$ for five images in Fig. \ref{fig:genifg}, which are slightly different outputs for the same input test interferogram of Fig. \ref{fig:qual_inphase}. Higher values of $\alpha$ generate more variations in the outputs, but also tend to make them noisy. This technique can be used in InSAR machine learning for data augmentation \cite{data_aug}, or to test any InSAR processing chain by running that chain all the way through, with slightly different interferograms to measure the variance of the outputs of the complete processing chain, and potentially, for error analysis.

\subsubsection{Experiments using simulated InSAR images}

\begin{figure*}
\centering

\subfigure[Input Phase]{\includegraphics[width=20.5mm]{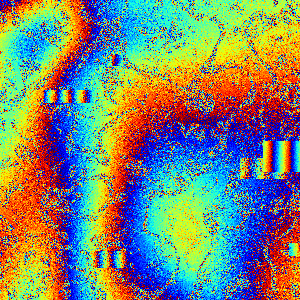}}
\enskip
\subfigure[Ground Truth Phase \& Coherence]{\includegraphics[width=20.5mm]{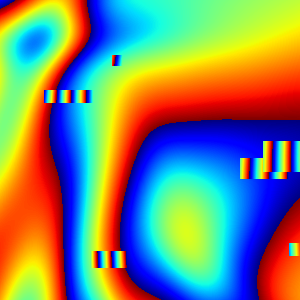}\enskip\includegraphics[width=20.5mm]{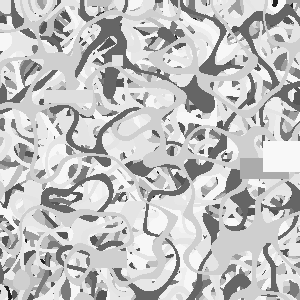}}
\enskip
\subfigure[Goldstein]{\includegraphics[width=20.5mm]{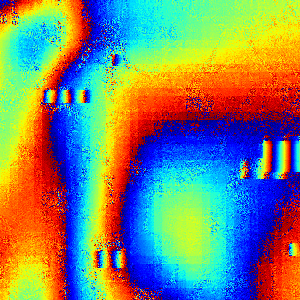}}
\enskip
\subfigure[Boxcar]{\includegraphics[width=20.5mm]{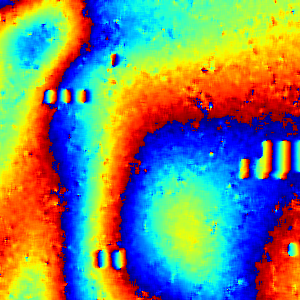}\enskip\includegraphics[width=20.5mm]{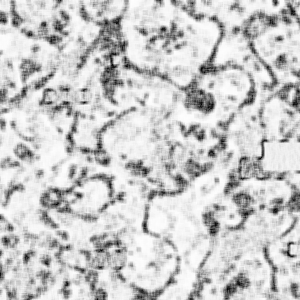}}
\enskip
\subfigure[NLInSAR]{\includegraphics[width=20.5mm]{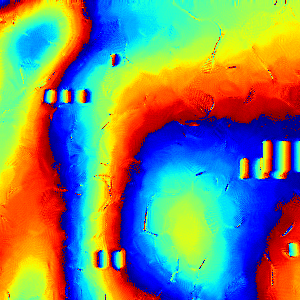}\enskip\includegraphics[width=20.5mm]{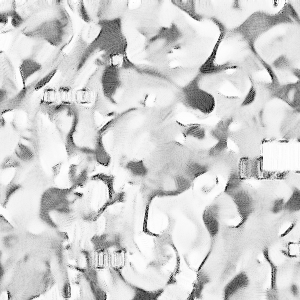}}
\enskip
\subfigure[NLSAR]{\includegraphics[width=20.5mm]{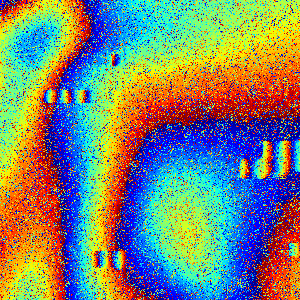}\enskip\includegraphics[width=20.5mm]{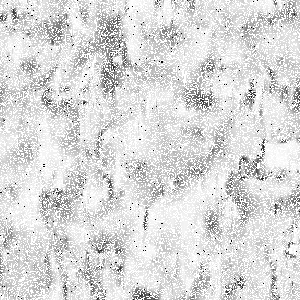}}
\enskip
\subfigure[CNN-InSAR (as-is)]{\includegraphics[width=20.5mm]{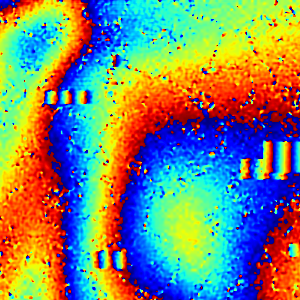}\enskip\includegraphics[width=20.5mm]{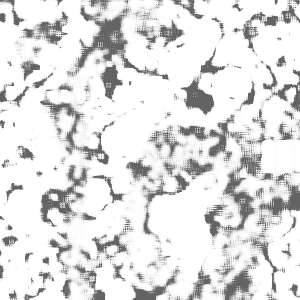}}
\enskip
\subfigure[CNN-InSAR (retrained)]{\includegraphics[width=20.5mm]{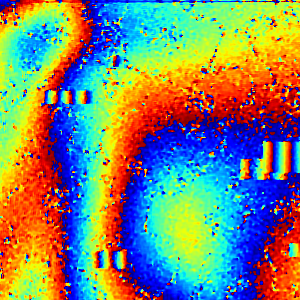}\enskip\includegraphics[width=20.5mm]{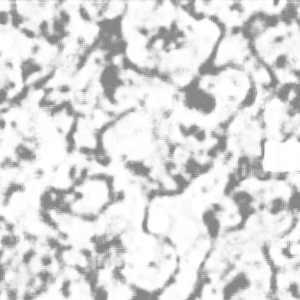}}
\enskip
\subfigure[Proposed Method (GenInSAR)]{\includegraphics[width=20.5mm]{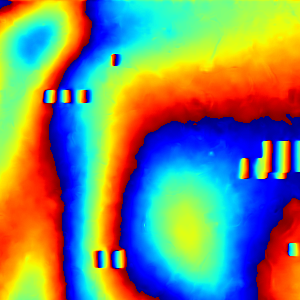}\enskip\includegraphics[width=20.5mm]{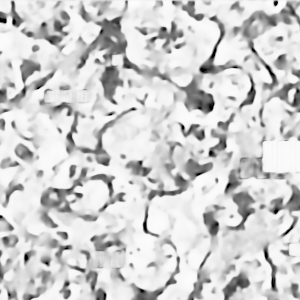}}

\caption{Filtered phase and coherence outputs for simulated InSAR images from GenInSAR (total residue reduction, less over-smoothing/artefacts around branch cuts) and five existing methods. Phase and coherence are coloured between --$\pi$ (blue) to +$\pi$ (red), and 0 (black: low) to 1 (white: high) respectively.}
\label{fig:squalcomp}
\end{figure*}

Our InSAR simulator can simulate ground truth interferograms with Gaussian bubbles, roads and buildings. We followed a similar training strategy as satellite InSAR images for training our model with simulated InSAR images, by adding Gaussian noise to simulated ground truth images, and inputting patches extracted from those noisy versions. For CNN-InSAR, we obtained two sets of results: 1. using the model as-is, and 2. retraining from scratch with simulated noisy images as mentioned above. For evaluating GenInSAR and five existing methods mentioned earlier including CNN-InSAR (as-is and retrained), we used 60 $1000\times1000$ pixels noisy simulated images. Fig. \ref{fig:squalcomp} compares all methods using a cropped region of one such test image. Corresponding full-size clean (ground truth) images facilitated quantitative evaluation (Table \ref{tab:results}) showing overall superior performance of GenInSAR against others and almost linear speedup with increasing number of GPUs.

GenInSAR almost totally reduces residues and produces far less over-smoothing/artefacts around branch cuts than Boxcar, because it's greatest strength is (unsupervised) learning of true spatial smoothing from noisy training data. It could potentially detect real residues better if trained more on such types of features, and a yet more efficient implementation, like the compared methods \cite{despeckCL} could further reduce it's run time. NLInSAR handles residues well and avoids artefacts by selecting neighbors with similar phase, but produces streaking correlated with amplitude bands. NLSAR (conservatively) interpolates well only over heavy noise. A final scope of future work for GenInSAR is enforcing the network's output pixel to lie on the unit circle, as currently, $(\sigma_{R}^2 + \sigma_{I}^2)$ is clipped to $[0,1]$, although even at present, most values lie in that range.

\begin{table}
\caption{Quantitative evaluation of GenInSAR and existing methods and scalability of GenInSAR over increasing GPU counts}
\label{tab:results}
\small
\setlength{\tabcolsep}{3pt}
\centering
\begin{tabular*}{\columnwidth}{@{\extracolsep{\fill}}|c|c|c|c|c|c|}
\hline
\makecell{Method \\ Name} & \makecell{Phase \\ RMSE} & \makecell{Coherence \\ RMSE} & \makecell{Residue \\ Red. \%} & \makecell{Cosine \\ Error ($\epsilon_{cos}^{\Delta\theta}$)} & \makecell{Time \\ (sec)} \\
\hline
\makecell{CNN-InSAR \\ (as-is)} & \makecell{1.270\\$\pm0.191$} & \makecell{0.257\\$\pm0.013$} & \makecell{92.74\\$\pm3.30$} & \makecell{0.060\\$\pm0.036$} & 1.42 \\
\makecell{CNN-InSAR \\ (retrained)} & \makecell{1.392\\$\pm0.192$} & \makecell{0.200\\$\pm0.025$} & \makecell{86.91\\$\pm4.71$} & \makecell{0.073\\$\pm0.040$} & 1.42 \\
NLSAR     & \makecell{1.537\\$\pm0.073$} & \makecell{0.301\\$\pm0.055$} & \makecell{35.85\\$\pm17.71$} & \makecell{0.132\\$\pm0.012$} & 11.49 \\
NLInSAR   & \makecell{0.850\\$\pm0.122$} & \makecell{0.159\\$\pm0.018$} & \makecell{97.59\\$\pm2.08$} & \makecell{0.014\\$\pm0.009$} & 20.44 \\
Goldstein & \makecell{1.260\\$\pm0.229$} & N/A & \makecell{88.51\\$\pm11.96$} & \makecell{0.048\\$\pm0.040$} & 2.17 \\
Boxcar    & \makecell{1.025\\$\pm0.173$} & \makecell{0.143\\$\pm0.018$} & \makecell{97.64\\$\pm1.94$} & \makecell{0.025\\$\pm0.021$} & 1.32 \\
\makecell{\textbf{Proposed}\\\textbf{(GenInSAR)}}  & \makecell{\textbf{0.687}\\$\pm0.102$} & \makecell{\textbf{0.138}\\$\pm0.015$} & \makecell{\textbf{99.78}\\$\pm0.08$} & \makecell{\textbf{0.005}\\$\pm0.004$} & \textbf{0.66} \\
\hline
\end{tabular*}
\begin{tabular*}{\columnwidth}{@{\extracolsep{\fill}}|c|c|c|c|c|c|c|c|}
GPU Count  & 64 & 32 & 16 & 8 & 4 & 2 & 1 \\
\hline
Time (sec) & \textbf{0.013} & \textbf{0.023} & \textbf{0.042} & \textbf{0.079} & \textbf{0.159} & \textbf{0.328} & \textbf{0.66} \\
\hline
\end{tabular*}
\end{table}

\begin{equation}
\label{eq:pce}
    \epsilon_{cos}^{\Delta\theta} = \frac{1}{n}\sum_{i=1}^{n}\frac{1}{2}(1-\cos(\arg{(g_{i}\bar{f_{i}})}))
\end{equation}


\section{Conclusion}

We propose an InSAR phase filter and coherence estimator that outperforms the state-of-art. Our generative modeling via unsupervised learning learns from petabytes of noisy InSAR data captured by an increasing number of SAR satellites, and can generate new interferograms to improve InSAR machine learning. Our GPU-based highly scalable filter can help monitor extensive areas of earth's surface for potential disasters.

\bibliography{bibliography}

\end{document}